\documentstyle[epsfig]{elsart}

\newcommand {\bi} {\bibitem}
 \newcommand {\be} {\begin{equation}}
\newcommand {\bea} {\begin{eqnarray} \nonumber }
\newcommand {\ee} {\end{equation}}
\newcommand {\eea} {\end{eqnarray}}
 
 \newcommand {\si} {\sigma}
\newcommand {\de} {\delta}

\newcommand {\ga} {\gamma}

 \newcommand {\al} {\alpha}

\newcommand {\lan} {\langle}
\newcommand {\ran} {\rangle}

\newcommand {\cH}  {{\cal H}}

\newcommand {\cN}  {{\cal N}}

\newcommand {\ap}  {\left(}
\newcommand {\cp}  {\right)}
\newcommand {\bc} {\begin{center}}
\newcommand {\ec} {\end{center}}
\newcommand {\bd}{\begin{displaymath}}
\newcommand {\ed}{\end{displaymath}}
\newcommand {\ato}  {\left(}
\newcommand {\cto}  {\right)}

\def \form#1 {eq. (\ref{#1}) }
\def \parziale#1#2  {{\partial {#1} \over \partial {#2}}}

\begin{document}
\begin{frontmatter}
    \title{The Physics of the Glass Transition}
\author{
Giorgio Parisi}
\address{Dipartimento di Fisica, Sezione INFN, Unit\`a INFM\\
Universit\`a di Roma ``La Sapienza'', 
Piazzale Aldo Moro 2,
I-00185 Rome (Italy)}

\begin{abstract}
\noindent In this talk, after a short phenomenological introduction on glasses, I will 
describe some recent progresses that have been done in glasses using the replica method in 
the definition and in the evaluation of the configurational entropy (or complexity).  
These results are at the basis of some analytic computations of the thermodynamic glass 
transition and of the properties below the phase transition point.
\end{abstract}
\end{frontmatter}
\section{Introduction}
Glasses are roughly speaking liquids that do not crystallize also at very low temperature 
(to be precise: glasses also do not quasi-crystallize).

These liquids can avoid crystalization mainly for two reasons:
\begin{itemize}
    \item The liquid does not crystallize because it is cooled very fast: the crystallisation time 
    may become very large at low temperature (e.g. hard spheres at high pressure).  The system 
    should be cooled very fast at temperatures near the melting point; however if crystalization is 
    avoided, and the temperature is low enough, (e.g. near the glass transition) the system may be 
    cooled very slowly without producing crystalization.
    
    \item The liquid does not crystallize even at equilibrium.  An example is a binary 
    mixture of hard spheres with different radius: 50\% type A (radius $r_{A}$, 50\% type 
    B (radius $r_{A}$, where $R$ denotes $r_{B}/r_{A}$.  If $ .77 <R <.89$ (the bounds may be not 
    precise), the amorphous packing is more dense than a periodic packing, distorted by 
    defects.
   
\end{itemize}

Which of the two mechanism is present is irrelevant for understanding the liquid glass
transition.

\begin{figure}
\centerline{\hbox{
\epsfig{figure=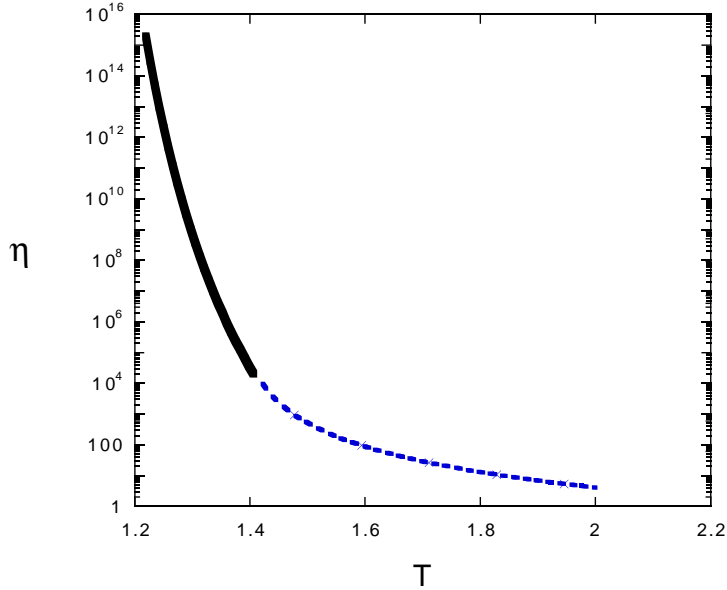,width=0.7\textwidth}
}}
\caption{The viscosity as function of the temperature according to the Vogel Fulcher law  
($T_{K}=1$ full line) and to the mode coupling theory  ($T_{K}=1.375$ dotted line).
 }
\label{fig_1}
\end{figure}

Other examples of glassy systems are mixtures: in this case we have $M$ kind of 
particles and the Hamiltonian of $N$ particles is given by
\be
H=\sum_{a,b=1,M}\sum_{i=1,N(a)}\sum_{k=1,N(b)} V_{a,b}\ato x_{a}(i)-x_{b}(k) \cto\ ,
\ee
where $a=1,M$; $N(a)=N c(a)$; $\sum_{a} c(a)=1$.  In this case the values of the $N$ 
concentrations $c$ and the $M(M+1)/2$ functions $V_{a,b}(x)$ describe the model.  Well 
studied case are:
\begin{itemize}
    
    \item Soft spheres (non-realistic, but simple), e.g.
$M=2$, $c(1)=.5$, $c(2)=.5$, $V_{a,b}(x) =R _{a,b} x^{-12}$.

\item 
Lennard-Jones spheres (more realistic), e.g.
$M=2$, $c(1)=.8$, $c(2)=.2$, $V_{a,b}(x) =R _{a,b} x^{-12} -A_{a,b}x^{-6}$.

\end{itemize}

There are some choices of the parameters which are well studied and in these case it is 
known that the system does not crystallize, one of these has been introduced by Kob and 
Anderson in the L-J case and correspond to a particular choice of the parameters 
$R$ and $A$.

There are many other material which are glass forming, e.g.
short polymers, asymmetric molecules (e.g. OTP) \ldots

The behaviour of the viscosity in glass forming liquid is very interesting.  We recall 
that the viscosity can be defined microscopically as follows.  We consider a system in a 
box of large volume $V$ and we call $T_{\mu,\nu}(t)$ the total stress tensor at time $t$.  
We define the correlation function of the stress tensor at different times:
\be
\lan T_{\mu,\nu}(t) T_{\rho,\si}(0)\ran = V S_{\mu,\nu,\rho,\si} (t)
\ee
One finds, neglecting indices, that
$\eta \propto \int dt S(t) \approx \tau^{-\al}$, where $\tau$ is the characteristic time of the 
system (in a first approximation we can suppose that
$\al=1$).

In fragile glasses the behaviour of the viscosity as function of the temperature can be 
phenomenological described by the two following regimes:
\begin{itemize}
    \item In an high temperature region the mode coupling theory is valid: it predicts 
    $\eta\propto (T-T_{c})^{-\ga}$, where $\ga$ is {\it not} an universal quantity and it 
    a of the orders of a few units.
    
    \item In the low temperature region by the Vogel Fulcher law is satisfied: it predicts that
    $\eta\propto \exp(A(T-T_{K})^{-1})$.
    
\end{itemize}
    
Nearly tautologically fragile glasses can be defined as those glasses that have $T_{K}\ne 
0$; strong glasses have $T_{K} \approx 0$.

The glass temperature ($T_{g}$) is the temperature at which the viscosity 
becomes so large that it cannot be any more measured. This happens after an increase 
of about 18 order of magnitude (which correspond to a microscopic time changing from 
$10^{-14}$ to $10^{4}$ seconds): the relaxation time becomes larger than the experimental 
tine.

A characteristic of glasses is the dependance of the specific heat on the cooling rate.  
There is a (slightly rounded) discontinuity in the specific heat that it is shifted at 
lower temperatures when we increase the cooling time.

\begin{figure}
    \epsfig{figure=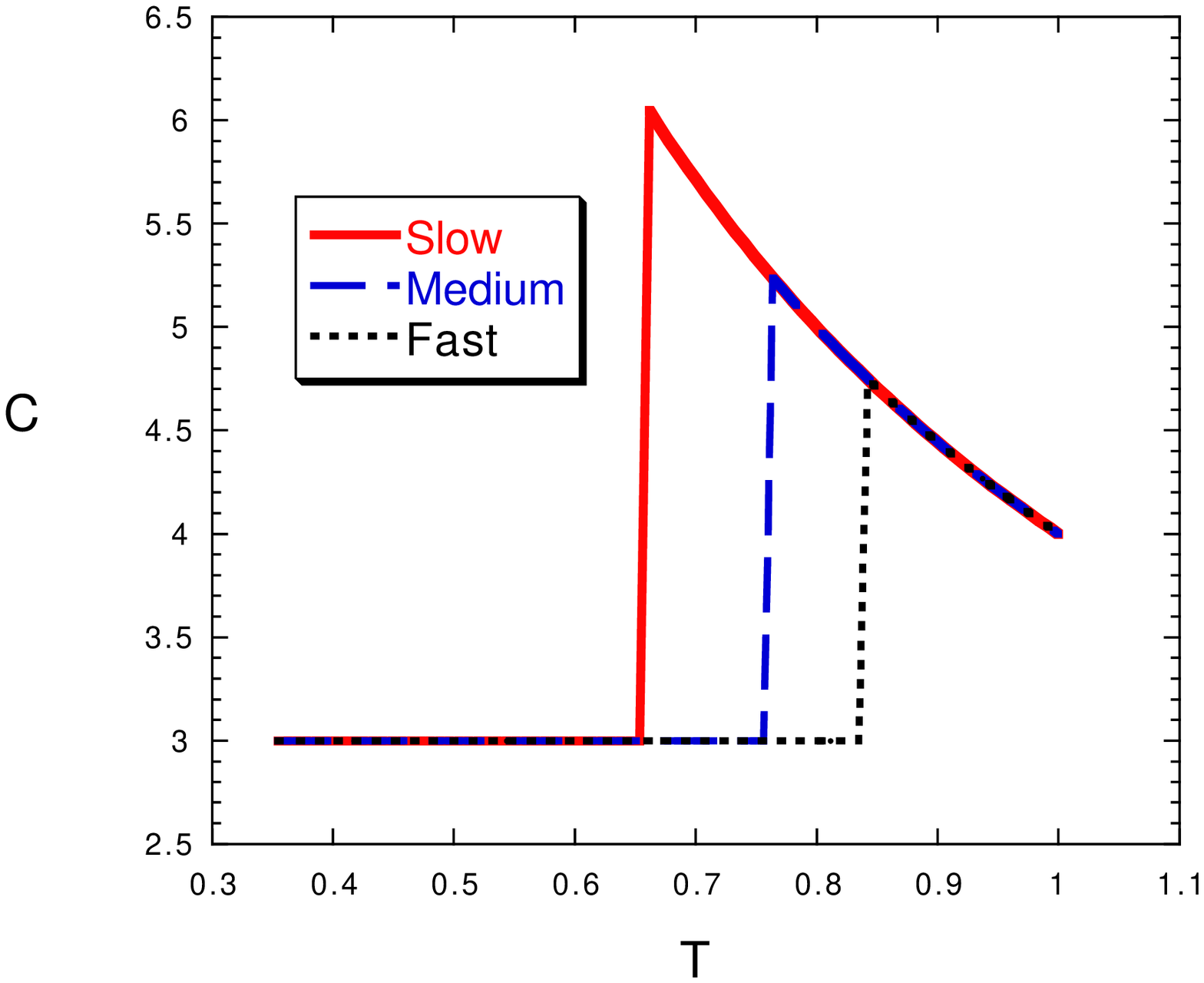,width=0.47\textwidth}
    \epsfig{figure=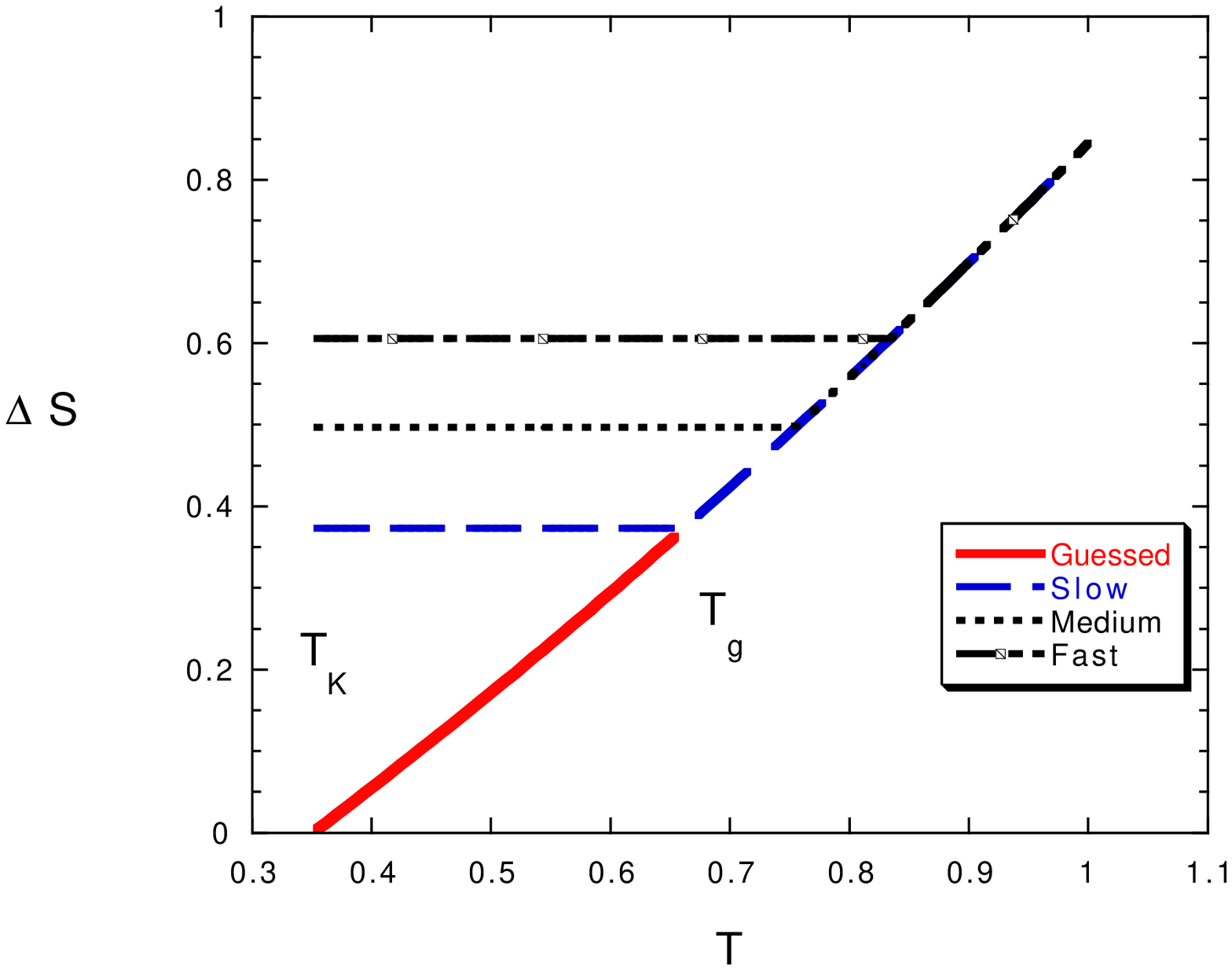,width=0.49\textwidth}
\caption{The specific heat and the entropy excess $\Delta S$ for various cooling rates.
 }
\label{fig_2}
\end{figure}

For systems that do crystallize if cooled very slowly, one can plot $\Delta S \equiv 
S(liquid) - S(crystal)$ versus $T$ in the thermalized region.  One gets a very smooth 
curve whose extrapolation becomes negative at a finite temperature.  A negative $ \Delta S 
$ does not make sense, so there is a wide spread belief that a phase transition is present 
before (and quite likely near) the point where the entropy becomes negative.  Such a 
thermodynamic transition (suggested by Kaufmann) would be characterized by a jump of the 
specific heat.  Quite often this temperature is very similar to the temperature found by 
fitting the viscosity by the Volker Fulcher law and the two temperature are believed to 
coincide.

Generally speaking it is expected that a thermodynamic phase transition induces a 
divergent correlation time.  However the exponential dependance of the correlation time on 
the temperature is not so common (in conventional critical slowing down we should have a 
power like behaviour); moreover an other strange property of the glass transition is the 
apparent absence of a detectable correlation length or susceptibility (linear or 
non-linear) which diverges when we approach the transition point.

In section \ref{MANY}, I will firstly present the many valley picture which has been 
the inspiration point of many works; in the subsequent section I will summarize some of the 
results that have been obtained using the replica theory on the organization of these 
valley in phase space, on the temperature dependence of the free energy in each valley, in 
brief of the {\sl free energy landscape}.  A strategy for applying these ideas to glasses 
will be presented in section \ref{MORE} and in the next section I will show some results 
that have been obtained in the case of a Lennard-Jones binary mixture.

\section{The many valleys picture}\label{MANY}

A quite old statement is that {\sl the glass is frozen liquid}, which I prefer to 
reformulate it as that {\sl a low temperature a liquid is an unfrozen glass}.
\begin{figure}
    \centerline{\hbox{
    \epsfig{figure=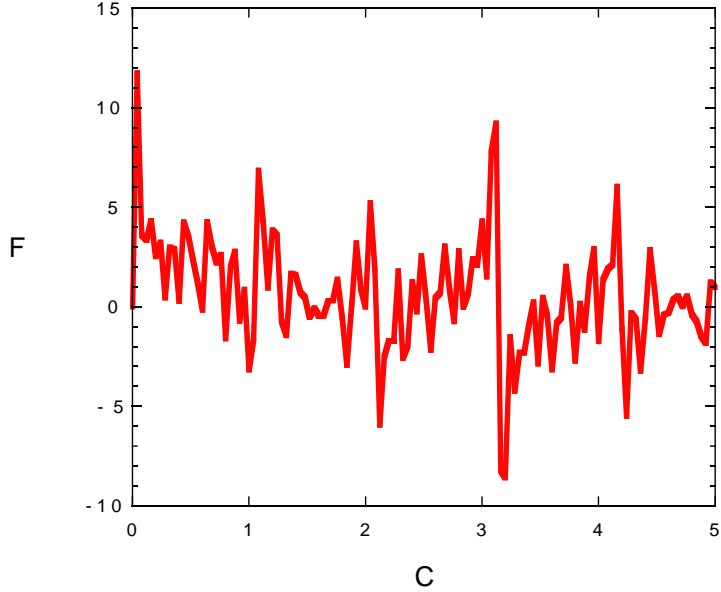,width=0.7\textwidth}
    }}
\caption{An artistic view of a corrugated free energy landscape, the horizontal axis being the 
configuration space.
 }
\label{fig_3}
\end{figure}

The general idea is the following: the phase space of the systems can be approximately 
decomposed into valleys, labelled by $a$, such that the barrier among these valleys is 
large in the liquid (in the low temperature region where the viscosity is large)   the 
barriers become infinite at $T_{K}$ \cite{VARI}.

For simplicity I will assume that valleys do maintain their identity when changing the temperature.  
(in reality they split into smaller valleys when we decrease the temperature \cite{BAFRPA}).  Under 
this assumption there is a one to one correspondence among valleys and inherent structures, i.e. 
minima of the Hamiltonian $H$.  The properties of the liquid are therefore connected to the 
properties of the system near the minima of the Hamiltonian.

The partition 
function can be approximately written as
\be
Z = \sum_{a}\exp \ato -\beta N f_{a}(\beta)\cto \ ,
\ee 
where $f_{a}(\beta)$ is the free energy density of the $a^{th}$ valley. As we shall see later 
in the infinite volume limit the 
previous sum is dominated by those valleys having a given free energy 
density $f=f^{*}(\beta)$.

The number of relevant valleys at $f=f^{*}(\beta)$ (i.e. $\cN^{*}(\beta)$) is supposed to 
exponentially increase with the size ($N$) of the system:
\be
cN^{*}\approx
\exp (N \Sigma^{*}(\beta))\ .
\ee
It can be argued that the configurational entropy or complexity
$\Sigma^{*}(\beta)$ is approximately near to $\Delta S$:
indeed it has suggest long time  ago that the configurational entropy $\Sigma^{*}$,
goes to zero at the phase transition point \cite{VARI}.

Recently using the techniques stemming from replica theory we have added new ingredients 
to this old scenario:

\begin{itemize}
\item
We have construct a microscopic realization of the previous ideas in the mean field models 
\cite{KiThWo,crisomtap}. 
In other words there are soluble infinite range models in which there is an exponentially 
large number of valleys and their properties can be computed analytically.
\item
If we assume that {\em minimal} corrections are present in finite dimensions to 
mean field predictions, the mean field results may be extended to three dimensional 
glasses.

\item The replica method \cite{mpv} gives tools to do the appropriate computations, both in mean 
the mean field approach and in short range models.

\end{itemize}

As outcome of these advances a many valleys picture with detailed 
predictions on the landscape  has been constructed and
the properties of glasses can be computed in the framework of replicated liquid theory 
\cite{MePa1,Me,sferesoft,LJ}.

\section{ The free energy landscape}

Let us suppose that we can define a temperature free energy functional ($F[\rho]$) as function of 
the density $\rho(x)$.  It is natural to assume that the different valleys are in a first 
approximation associated to different local minima of this functional, the free energy evaluated at 
the local minima being the free energy of the corresponding valley.

I will summarize which are in mean field the properties of the local minima the free energy that 
have been computed explicitly in some long range microscopic models.  Although these results have 
been obtained in some particular models, one can show that they are valid for a quite large class of 
models (at least in the mean field approximation \cite{FP,CFP}).
\begin{itemize}
    \item For $T>T_{c}$ there is one relevant minimum with $\rho(x)=const$.  There are 
    many solutions of the equation 
    \be {\de F\over \de \rho}=0.
\ee
which are not minima (there are some negative eigenvalues of the Hessian $\cH(x,y)
={\partial^{2} H \over \partial x \partial y}$.

\item For $T=T_{c}$ the negative eigenvalues of the Hessian become positive or zero.  
There is an exponentially large number of minima, connected by flat regions.

\item For $T_{c}>T>T_{K}$ the number of relevant minima becomes proportional to 
$\exp( N \Sigma^{*}(\beta))$. 
where
$\Sigma^{*}(\beta)$ is the configurational entropy or complexity.  
The minima are separated by barriers that
diverge with the number of particles $N$ (in mean field theory), however the barriers are
are finite in real life (i.e. beyond mean field theory).

\item For $T_{K}>T$ the number of relevant minima is finite.  
The minima are separated by 
barriers that
diverge with $N$.
\end{itemize}

For $T_{c}>T>T_{K}$ there is a dual description: the system may be described as a {\em 
liquid} and  also as a {\em solid} (with an exponentially large number of different 
solid structures).  This duality tell us that all the information needed to study the 
glass can be extracted from the liquid phase.

\section{A more quantitative approach}\label{MORE}

As we have seen we can write
\be
Z(\beta)=
\sum_{a} \exp( -\beta N f_{a}(\beta))= 
\int d \cN(f,\beta) \exp (-\beta N f),
\ee
where $f_{a}(\beta)$ is the free energy density of the valley 
labeled by $a$ at the temperature $\beta^{-1}$, and
$\cN(f,\beta)$ is the number of valleys with free energy density  less than 
$f$.

We can assume that $\cN(f,\beta) =\exp(N \Sigma(f,\beta))$, where the configurational 
entropy, or complexity, $\Sigma(f,\beta)$ is positive in the region $f.f_{0}(\beta)$ and 
vanishes at $f=f_{0}(\beta)$.  The quantity $ f_{0}(\beta)$ is the minimum value of the 
free energy: $\cN(f,\beta)$ is zero for $f< f_{0}(\beta)$ \cite{MePa1,Me,sferesoft,LJ}.

If the equation
\be
{\partial \Sigma \over \partial f} =\beta
\ee
has a solution at $f=f^{*}(\beta)$ (obviously this may happens only for 
$f^{*}(\beta)>f_{0}(\beta)$), we stay in the liquid phase the free density is given by
\be
F=f^{*}-\beta^{-1} \Sigma(f^{*},\beta)
\ee
and $\Sigma(f^{*},\beta)= \Sigma^{*}(\beta)$.

Otherwise we stay in the glass phase and 
\be
F=f_{0}(\beta)\ .
\ee

In order to compute the properties in the glass phase we 
{ need} to know $\Sigma(f,\beta)$: a simple strategy is the following.

We introduce the modified partition function
\be
Z(\gamma;\beta)\equiv \exp ( -N \gamma \Phi(\gamma;\beta))=\sum_{a} \exp( -\gamma N 
f_{a}(\beta)).
\ee
Using standard thermodynamical arguments it can be easily proven that in the limit $N 
\to \infty$ one has:
\bea
\gamma \Phi(\gamma;\beta)=  \gamma f - \Sigma(\beta,f),\ \ \ 
f={\partial (\gamma \Phi(\gamma;\beta) )\over \partial \gamma}.
\eea

The complexity is obtained from $\Phi(\gamma;\beta)$ in the same way as the 
entropy is obtained from the usual free energy \cite{MONA}:
\be
\Sigma(\beta,f)={\partial \Phi(\gamma;\beta) \over \partial \gamma}.
\ee

A few observations are in order:
\begin{itemize}
\item In the new formalism $\gamma$, the free energy and the complexity play respectively 
the same role of $\beta$, the internal energy and the entropy in the usual formalism.  

\item In the new formalism $\beta$ only indicates the value of the temperature which is 
used to compute the free energy and $\gamma$ controls which part of the free energy 
landscape  is sampled.  

\item When $\beta \to \infty$ we sample the energy landscape: 
\be
Z(\gamma;\infty)=\sum_{a} \exp( -\gamma N e_{a})=\int \nu(e) de \exp (- \gamma N e)
\ee
where $e_{a}$ are the minima of 
the Hamiltonian and $\nu (e)$ the density of the minima of the Hamiltonian.

\item The equilibrium complexity is given by 
$\Sigma ^{*}(\beta) = \Sigma (\beta;\beta)$. On the other hand $\Sigma (\gamma;\infty)$ 
give us information on the minima of the Hamiltonian.
\end{itemize}

It is convenient to write 
\bea
Z(\gamma;\beta)=
\int  d C \exp \ato-\gamma H(C) -N \gamma  f(\beta,C) + N\gamma 
f(\gamma,C)\cto=\\
\int  d C \exp \ato-\gamma H(C) -N \gamma  \hat{f}(\beta,C) + N \gamma 
\hat{f}(\gamma,C)\cto,
\eea
where  
$\hat{f}(\beta,C)= f(\beta,C)-f(\infty,C)$,
$f(\beta,C)$ is is constant in each valley and it is  equal to the free 
 energy density
of the valley to which the configuration $C$ belongs.
and $\hat{f}(\beta,C)$ is the temperature dependent part of free 
 energy density
of the valley to which the configuration $C$ belongs.

Our strategy is to use duality and liquid theory to evaluate the properties of 
$\hat{f}(\beta,C)$.

A simple approximation is the so called quenched approximation which in this case 
consists in assuming that
\be
\exp (-A \hat{f}(\beta,C))\approx\exp(-A \hat{f}_{\gamma}(\beta)),
\ee
where $\hat{f}_{\gamma}(\beta)$ is the expectation value of 
$\hat{f}(\beta,C)$ taken with the probability 
distribution proportional to $\exp(-\gamma H(C))$. 

The quenched approximation would be exact if all the valleys at a given 
temperature (also with different free energy) would 
have the same entropy. In other words the fluctuations of the energy are more relevant 
than the fluctuations of the entropy.

It is now a matter of simple algebra to show that
\bea
\Phi(\gamma;\beta)= F_{liquid}(\gamma)+ \hat{f}_{\gamma}(\beta)  
-\hat{f}_\gamma(\gamma),\\
\Sigma(\gamma;\beta)=
S_{liquid}(\gamma)- S_{\gamma}(\gamma) +\hat{f}'_{\gamma}(\gamma) -
  \hat{f}'_{\gamma}(\beta), \label{FINAL}
\eea
where
$
\hat{f}'_{\gamma}(\beta) = {\partial \hat{f}_{\gamma}(\beta) / \partial \gamma}
$.

Everything is now computable using liquid theory.  Different approximations can be done 
for $\hat{f}_{\gamma}(\beta)$ and slightly different results may be obtained for the 
transition temperature and for the specific heat in the glass region.  The replica 
formalism allow us to study directly the function $\Phi(\gamma;\beta)$ and most of the 
computation are done (for sake of compactness and simplicity) using this powerful tool.

\section{Some results}

Analytic and numerical computation have been done in many case. I will present here only 
the results for the Kob Andersen Lennard Jones binary mixture described in the 
introduction at density $\rho=1.2$.

In order to compute the free energy (or the entropy of the valleys) a simple and useful 
approximation 
is to assume that the potential near each minimum of the Hamiltonian is approximately 
an harmonic one and therefore the free energy of a valley can be computed using an 
harmonic approximation.  In this way one finds that the previous formulae in the 
liquid phase become
\be
\Sigma^{*}(\beta))=S_{liquid}(\beta)-S_{harmonic}(\beta)+\ldots,
\ee
where
\be
\beta S_{harmonic}= \frac12 \ln \det \ap{\partial^{2} \beta H\over \partial x(i) \partial x(k)}
\cp.
\ee

Similar formulae holds below the transition temperature.

The relevant quantities can be or computed analytically (using liquid theory) or extracted 
by numerical simulations in the liquid phase.
\begin{figure}
\centerline{\hbox{
\epsfig{figure=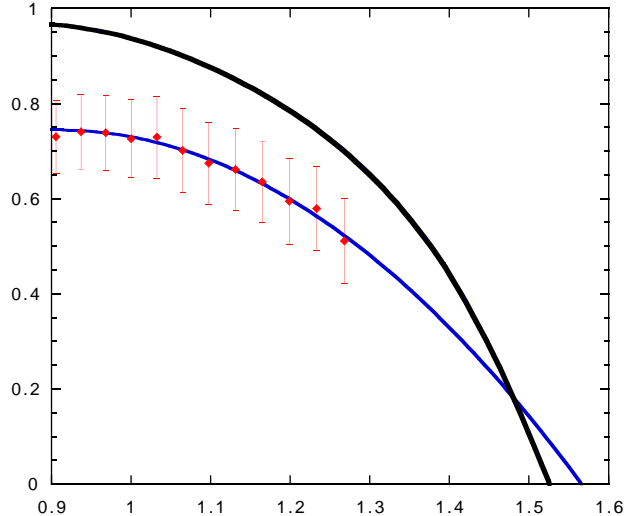,width=0.6\textwidth}
}}
\caption{Analytic and numerical complexity as function of $T^{-.4}$. }
\label{fig_Sc}
\end{figure}

Both the analytic and the numerical results of \cite{LJ} for the configurational are shown in 
fig. \ref{fig_Sc} (from \cite{LJ}). The correctness of the numerical estimate have been later confirmed by 
the more accurate computations of \cite{KST}.

These microscopic analytic computations support the proposal that there is a thermodynamic 
liquid glass transition characterized by vanishing of the complexity and that 
at low temperature below $T_{K}$ the system stays only in a few valleys.

In other words in the glass phase replica symmetry is broken and we aspect that there are 
characteristic violations of the fluctuation dissipation theorem in off-equilibrium dynamics in the 
aging regime \cite{CuKu1,FRAPA,FrMePaPe}.  Indeed in the aging regime the fluctuation dissipation theorem cannot be applied and 
generalized dissipation relations are valid.  The form of these generalized dissipation relations is 
fixed by the theory and the resulting predictions are confirmed by numerical simulations.

Direct measurements of these fluctuation dissipation relations in real (not numerical \cite{FDT}) 
experiments both for glasses are possible with the present technology.  Some experiments 
are under the way.  The outcome will be a crucial test for this theoretical approach and I 
hope that some results will be available in the next future.

\end{document}